\documentclass{aa}  

\usepackage{graphicx}
\usepackage{txfonts}
\usepackage{verbatim}
\newcommand{\twco}{{\hbox {\ensuremath{\mathrm{^{12}CO}} }}}
\newcommand{\thco}{{\hbox {\ensuremath{\mathrm{^{13}CO}} }}}
\newcommand{\kmps}{\ensuremath{\mathrm{km\,s^{-1}}}}

\newcommand{\Msun}{\ensuremath{{M}_\odot}}

\newcommand{\Tmb}{\ensuremath{{T}_\mathrm{MB}}}

\newcommand{\tastar}{\hbox {{\it T}$_{\rm A}^*$}}

\newcommand{\HII}{\ion{H}{II}}

\begin{document}

\title{ALMA detection of the dusty object silhouetted against the S0
  galaxy NGC\,3269 in the Antlia cluster\thanks{Molecular line
    spectra (Spectra extracted from a spatially smoothed ALMA image as
    a FITS binary table) are only available in electronic form at the CDS via
    anonymous ftp to cdsarc.u-strasbg.fr (130.79.128.5) or via
    http://cdsweb.u-strasbg.fr/cgi-bin/qcat?J/A+A/}}

    \author{L.\,K.  Haikala          \inst{1}              
\and
            R. Salinas               \inst{2}
\and
            T. Richtler              \inst{3}
\and
            M. G\'omez                 \inst{4}
\and          
            G.\,F. Gahm                  \inst{5}
\and          
            K. Mattila               \inst{6}
    }
 
   \institute {Instituto de Astronom\'ia y Ciencias Planetarias de Atacama,
               Universidad de Atacama, Copayapu 485, Copiapo, Chile  \\
              \email{lauri.haikala@uda.cl}
\and
               Gemini Observatory/NSF's NOIRLab, Casilla 603, La Serena, Chile
\and
               Departamento de Astronom\'ia, Universidad de Concepci\'on,
               Concepci\'on, Chile
\and
               Departamento de Ciencias F\'isicas, Facultad de Ciencias Exactas,
               Universidad Andres Bello, Fern\'andez Concha 700, Las Condes\\
               7591538, Chile
\and
               Stockholm Observatory, AlbaNova University Centre,
               Stockholm University, 106 91 Stockholm, Sweden
\and
               Department of Physics, University of Helsinki, P.O. Box 64,
               FI-00014 Helsinki, Finland 
   }

   \date{Received }

 
  \abstract
  {An intriguing silhouette of a small dust patch can
    be seen against the disk of the S0 galaxy
    NGC\,3269 in the Antlia cluster in optical images. The images do not
    provide any clue as to whether the patch is a local Jupiter mass-scale cloudlet
  or a large extragalactic dust complex.}
    {We aim to resolve the nature of this object: is it a small Galactic cloudlet 
    or an  extragalactic dust complex?}
    {ALMA and APEX  spectroscopy and
      Gemini GMOS long-slit spectroscopy were used to measure the velocity of
      the patch and the NGC\,3269 disk radial velocity curve.}
    {A weak 16$\pm 2.5$\,\kmps\ wide \element
    [][12]{C}O (2--1) \Tmb\ 19$\pm 2.5.$\,mK line in a 2\farcs2 by
    2\farcs12 beam associated with the object was detected with
    ALMA. The observed heliocentric velocity, $V_{\rm r,hel} =
    3878\pm5.0$\kmps,\ immediately establishes the extragalactic
    nature of the object. The patch velocity is consistent with the
    velocity of the nucleus of NGC\,3269, but not with the radial
    velocity of the NGC\,3269 disk of the galaxy at its position.  The
    $\sim$4\,\arcsec\ angular size of the patch corresponds to a linear
    size of $\sim$1 kpc at the galaxy's Hubble distance of 50.7\,Mpc.
    The mass estimated from the \element [][12]{C}O (2--1) emission is
    $\sim$1.4$\times 10^6 (d/50.7\,\mathrm{Mpc})^2$\,\Msun, while the
    attenuation derived from the optical spectrum implies a dust mass
    of $\sim $2.6$\times 10^4 (d/50.7\,\mathrm{Mpc})^2$\,\Msun.  {The
      derived attenuation ratio $A'_B$/($A'_B-A'_R$) of $1.6\pm$0.11
      is substantially lower than the corresponding value for the mean
      Milky Way extinction curve for point sources (2.3). } }
      {We established the extragalactic nature of the patch, but its
        origin remains elusive. One possibility is that the dust patch
        is left over from the removal of interstellar matter in
        NGC\,3269 through the interaction with its neighbour,
        NGC\,3268.}


\authorrunning{Haikala et al. }
\titlerunning {Dusty object silhouetted against NGC\,3269}
   \maketitle
%

\section{Introduction} \label{intro}

\begin{figure}
\centering
\includegraphics[width=88mm]{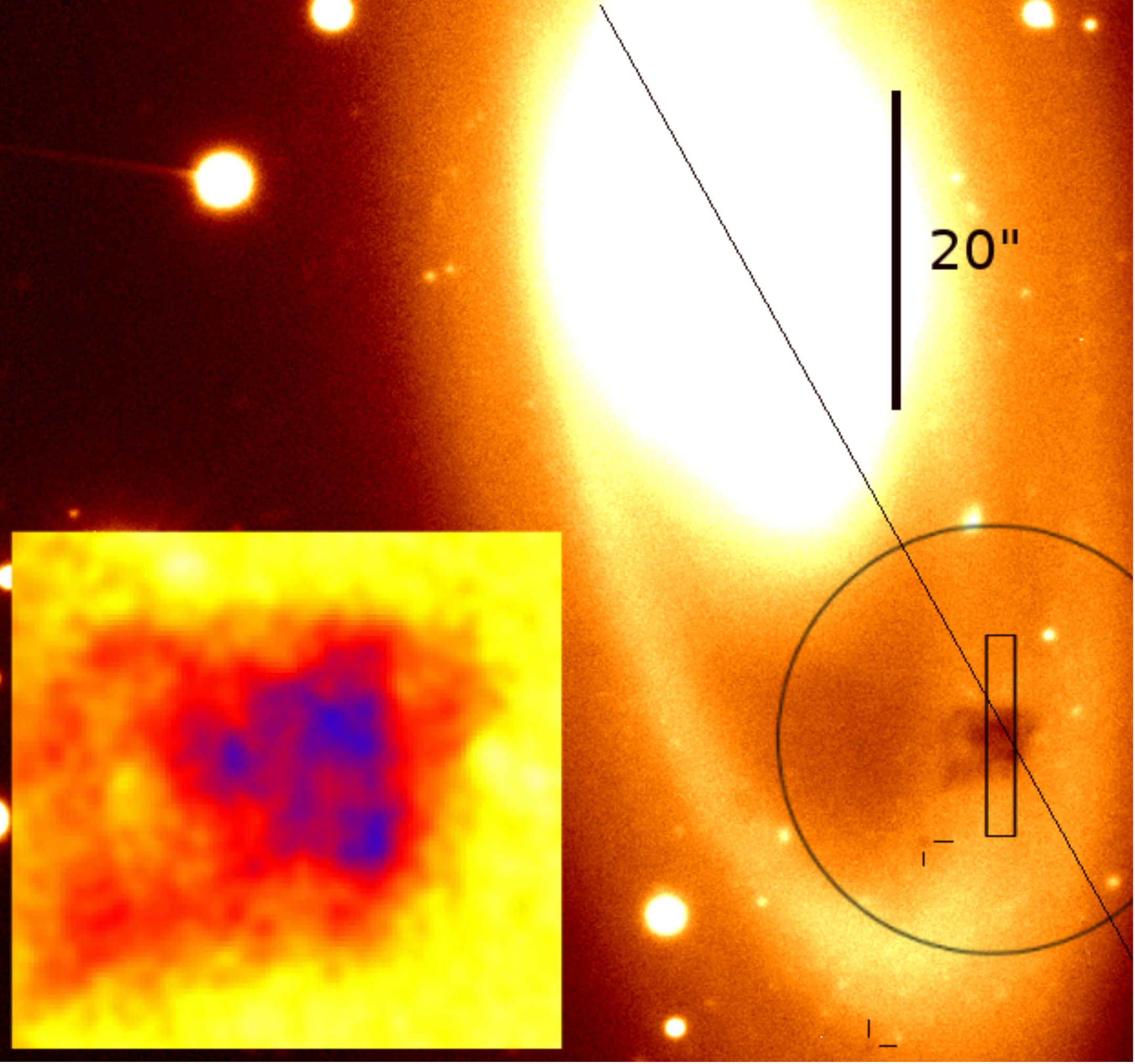}
\caption{Detail of Magellan telescope IMACS Sloan $r$-band image
  of NGC\,3269. North is up and east to the left. {\emph{The patch}} lies
  34\arcsec\ south-west of the bright nucleus. The line indicates the
  orientation of the GMOS long-slit through the nucleus and {\emph{the
    patch}}.  The two tiny patches and the virtual slit used in \citet
  {dirschetal_2005} are indicated. The circle shows the full width at half maximum (FWHM) of
    the ALMA primary beam at 228\,GHz. The 6\arcsec\ by 6\arcsec
    inset in the lower left shows a smoothed contrast-enhanced  Magellan Instant Camera
    (MagIC) $B$\ -band image \citep {dirschetal_2005} of {\emph{the
    patch}}.}
  \label{Magellan1}%
 \end{figure}

 The projection of a Galactic dust feature onto a distant galaxy is
 very rare. To our knowledge, only \citet
 {dirschetal_2003,dirschetal_2005} explicitly consider such a case.
 They noted a tiny dust patch with a diameter of about
 4\arcsec\ (hereafter referred to as {\emph{'the patch'}}) and two
 nearby smaller 0\farcs5 diameter patches seen as silhouettes against
 the S0 galaxy \object {NGC\,3269} located in the area of the Antlia
 cluster.  \citet {dirschetal_2005} estimated the B-band attenuation
 caused by {\emph{the patch}} to be approximately one magnitude. A
 recent high-quality Sloan $r$-band image, taken with Inamori-Magellan
 Areal Camera \& Spectrograph (IMACS) at the
 Magellan Baade Telescope, shows the structure of {\emph{the patch}}
 more clearly
 than before (Fig.\,\ref{Magellan1}).  The smoothed and
 contrast-enhanced inset in Fig.\,\ref{Magellan1} reveals that
 {\emph{the patch}} consists of three arcsecond-scale attenuation
 maxima surrounded by a less dense cometary-shaped halo. The optical
 images provide only an upper limit for the distance of {\emph{the
     patch}}, that of NGC\,3269, and thus the linear size and nature
 of the object is unknown.

Sub-arcsecond to 20\arcsec\ dusty objects, such as globulettes,
residing in \HII\ regions are detected optically \citep
[e.g.][]{reipurthetal2003,demarcoetal2006,gahm07} and even in the NIR
\citep [e.g.][]{makela14}, as they are seen as silhouettes against the
bright nebular background emission.  Despite their small sizes
(4\arcsec\ to 20\arcsec), the globulettes in the Rosette and Carina
nebulae were well detected in \twco and \thco (3--2) and (2--1) lines
at Atacama pathfinder experiment (APEX) \citep
{gahm13,Haikalaetal2017}.  It has been suggested \citep {lawrence2001}
that some of the common unidentified faint sub-millimetre sources found in
recent surveys are small cold (7\,K) clouds of dust and gas with
masses of $\sim$10$^{-4} - 10^{-2}$ \Msun\ (0.1 -- 10 Jupiter
masses) residing in our neighbourhood, $r < 100$\,pc. In a blind
single-dish CO\,(1-0) emission-line search, followed up by
high-angular resolution ($\sim$3\arcsec) interferometry,
\citet{heithausen2002,heithausen2004,heithausen2006} detected
cloudlets with sizes of a few hundred AU if at the adopted distance of
100 pc, suggesting that such clumpuscules may be an abundant
phenomenon in the local interstellar space. Detecting the clumpuscules
optically is practically impossible, because if they are not
associated with \ion{H}{ii} regions, they lack bright background
emission. Even if detected, the optical images would not provide any
clues on the distance to clumpuscules unless they were associated with
a known source.

{\emph{The patch}} could be a tiny, nearby 400 AU diameter sub-Jupiter mass
clumpuscule in the Milky Way at a distance of $\sim$100 pc. This would
be the first optical detection of a clumpuscule. The association with
NGC\,3269 would also be intriguing. NGC\,3269 is a S0 galaxy with a
grand design spiral pattern that is still visible (Fig.\,\ref{Magellan1}), and
with a heliocentric radial velocity of 3750 to 3797\,\kmps. There are no
signs of actual or recent star formation. The existence of an isolated
dust complex far from the centre is also very atypical for the
appearance of dust in early-type galaxies.  Moreover, the galaxy seems
to be devoid of \ion {H}{I} \citep{barneswebster2001}, with an upper
limit of about $10^9$\,\Msun.  If associated with NGC\,3269, {\emph{the
  patch}} would be an extragalactic giant gas/dust cloud complex larger
than 500\,pc in diameter.

  We utilised the APEX telescope and followed up with the ALMA
  12m array to search for CO emission in the direction of {\emph{the
    patch}}.  Detection of CO emission from {\emph{the patch}} at Galactic
  velocities around zero would provide the first freely floating sub-Jupiter mass
  object that can be investigated in detail both in
  radio and optical. On the other hand, a  radial velocity near that of
  NGC\,3269 would confirm its extragalactic nature.
  In addition, we obtained an optical spectrum with
  the Gemini Multi-Object Spectrograph (GMOS) at Gemini South to measure the
  radial velocity of the NGC\,3269 disk at the position of {\emph{the
    patch}}.  Observations and data reduction are described in
  Sect. \ref{obs}.  The results are presented in Sect. \ref{res} and
  discussed in Sect. \ref{discuss}, and the conclusions are summarised
  in Sect. \ref{conclusions}.

\section{Observations and data reduction} \label{obs}
\subsection{APEX} \label{obsapex}
Observations (project O-079F-9321A, PI Gahm) of the \twco J\,=\,(3--2) transition
at 345.796 GHz were carried out on September 27, 2007 in good
weather (PWV 0.34 mm) with the 12\,m APEX telescope at Llano
Chajnantor, Chile.  We used the double-side-band (DSB) heterodyne
SIS-receiver APEX-2A mounted on the APEX Nasmyth-A focus.  All
observations were performed in position-switching mode. The telescope
FWHM is 18\arcsec\ at 345 GHz, and the main-beam efficiency is
0.73. The pointing is estimated to be within
2\arcsec.  The FFTS1 spectrometer had a bandwidth of 1\,GHz and a channel
width of 0.4883\,MHz, corresponding to a channel width
of 0.423\,\kmps\ at 345\,GHz.  Calibration was achieved by the chopper
wheel method. The difference in the atmospheric opacities in the two
side-bands was estimated using an atmospheric model and was taken into
account in the calibration.  Typical values for the effective DSB
system temperatures outside the atmosphere were around 170\,K.

Two spectra, each covering $\sim$870\,\kmps\ in velocity, one centred at
0\,\kmps\ and another at 3799\,\kmps, were obtained. The total on-source
integration times were 12 minutes and 51 minutes,
respectively. The first-order baseline was subtracted, and the resulting
spectrum rms in \tastar\ scale was 0.04\,K (0.029\,K after Hanning-smooth)
for the spectrum centred at 0.0\,\kmps. For the second spectrum,
the rms was 0.02\,K (0.015\,K).

\subsection{ALMA 12m} \label{obsalma}

 {\emph{The patch}} was observed during  ALMA Cycle 5
 (Project 2017.1.00066.S) in the band-6  \twco (2--1) line at
 230.538\,GHz. Three spectral bands covered all the velocities from 
 -150\,\kmps\ up to 4075\,\kmps,\ and one band was dedicated to
 continuum. Details of the observations are given in Table \ref{t1}.
 The observations were conducted in good weather conditions (PWV
 0.677\,mm or 2.75\%) on August 16,$^{}$\ 2018 using 44 antennas in
 configuration C43-2 (baselines 15\,m to 500\,m, maximum resolvable scale
 7\farcs8).  The field was centred at 10$^h$29$^m$55\farcs8,
 -35\degr13\arcmin58\farcs0 J2000.0, and the achieved spatial
 resolution was 0\farcs73 by 0\farcs96, position angle -80\degr.

\begin{table}
\caption{\label{t1} ALMA Cycle 5 \twco (2--1) observations} \centering
\begin{tabular}{lcccccccccc}
\hline\hline
Band & $\nu_{cen}$   & Width & $\Delta\nu$ & $\Delta$v &  rms  \\
 &  [GHz] & [GHz]& [MHz]  & [\kmps] & [mK] \\
\hline
Galactic            & 230.414  &  468.8 &   0.244   &   0.318    &     2.0   \\
Intergalactic       & 229.726  &  938   &   0.488   &   0.635    &     1.3   \\
Extragalactic       & 228.327  &  1875  &   0.977   &   1.28     &     1.0   \\
\hline
\end{tabular}
\end{table}

\subsection{Gemini-S spectroscopy}

The galaxy NGC\,3269 was observed with the Gemini South telescope, located in
Cerro Pach\'on, Chile, using GMOS
\citep{hook04}, on the nights of February 4 and 5, 2019, under programme
GS-2018B-FT-208. GMOS is equipped with three Hamamatsu CCDs
\citep{gimeno16}, for a field-of-view  of 5.5\arcmin$\times$5.5\arcmin. A
1\arcsec\ long-slit together with the B600 grating gave a 4.6\AA
\ spectral resolution. The slit was positioned {and aligned to
  include both the galaxy's optical centre and {\emph{the patch}}}. In
total, ten exposures of 1200\,s each were obtained, together with
spectroscopic flats and CuAr arcs obtained at similar elevation.

Data reduction was conducted with standard techniques implemented
within the Gemini IRAF package. The  wavelength zero point  was corrected
by measuring the position of the bright sky line \ion{O}{I}\,6300.30\AA
\  after bias subtraction, flat-fielding and wavelength calibration.
Two-dimensional spectra were then combined with the task \verb+gemscombine,+ and sky
was subtracted with a sky sample $\sim$75\arcsec\  away from the galaxy
centre, where no galaxy signal is visible. Spectra were finally
extracted at different radii with \verb+apall+.

\subsection{Magellan IMACS imaging.}

{The galaxy NGC\,3269 was observed with the Magellan IMACS f/4 8Kx8K CCD
  mosaic camera (pixel size 0\farcs11) on the night of February 10,
  2018.  3$\times$600\,s, 2$\times$600\,s and 2$\times$600\,s exposures
  were taken in Sloan $g$, $r$, and $i$, respectively, with an average seeing
  of 0.45\arcsec. Images were reduced with THELI \citep{schirmer13},
  including bias subtraction, flat-fielding, mosaicing of the eight CCDs,
  and final co-addition.}

\begin{figure}[tbh]
\includegraphics[width=8.8cm]{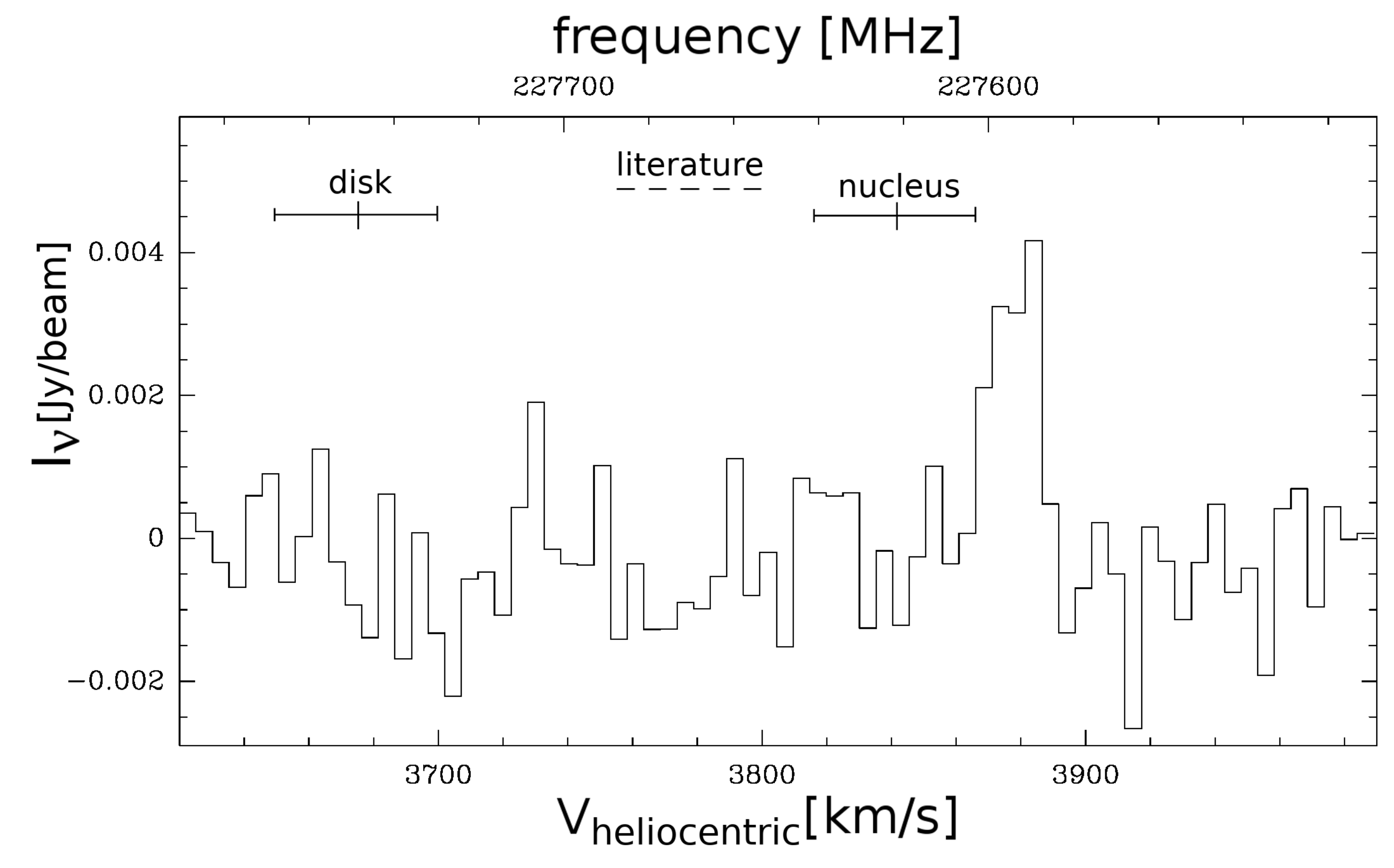}
\caption {Twice-Hanning-smoothed ALMA \twco (2--1) spectrum in a
  2\farcs2 by 2\farcs12 beam towards {\emph{the patch}}. The heliocentric
  radial velocity of the NGC3269 disk and nucleus as measured from the
  GMOS spectrum and the range of NGC3269 velocity values in the
  literature are indicated.  }
\label{Patchspec}
\end{figure}

 \begin{figure}
\centering
\includegraphics[width=8.8cm]{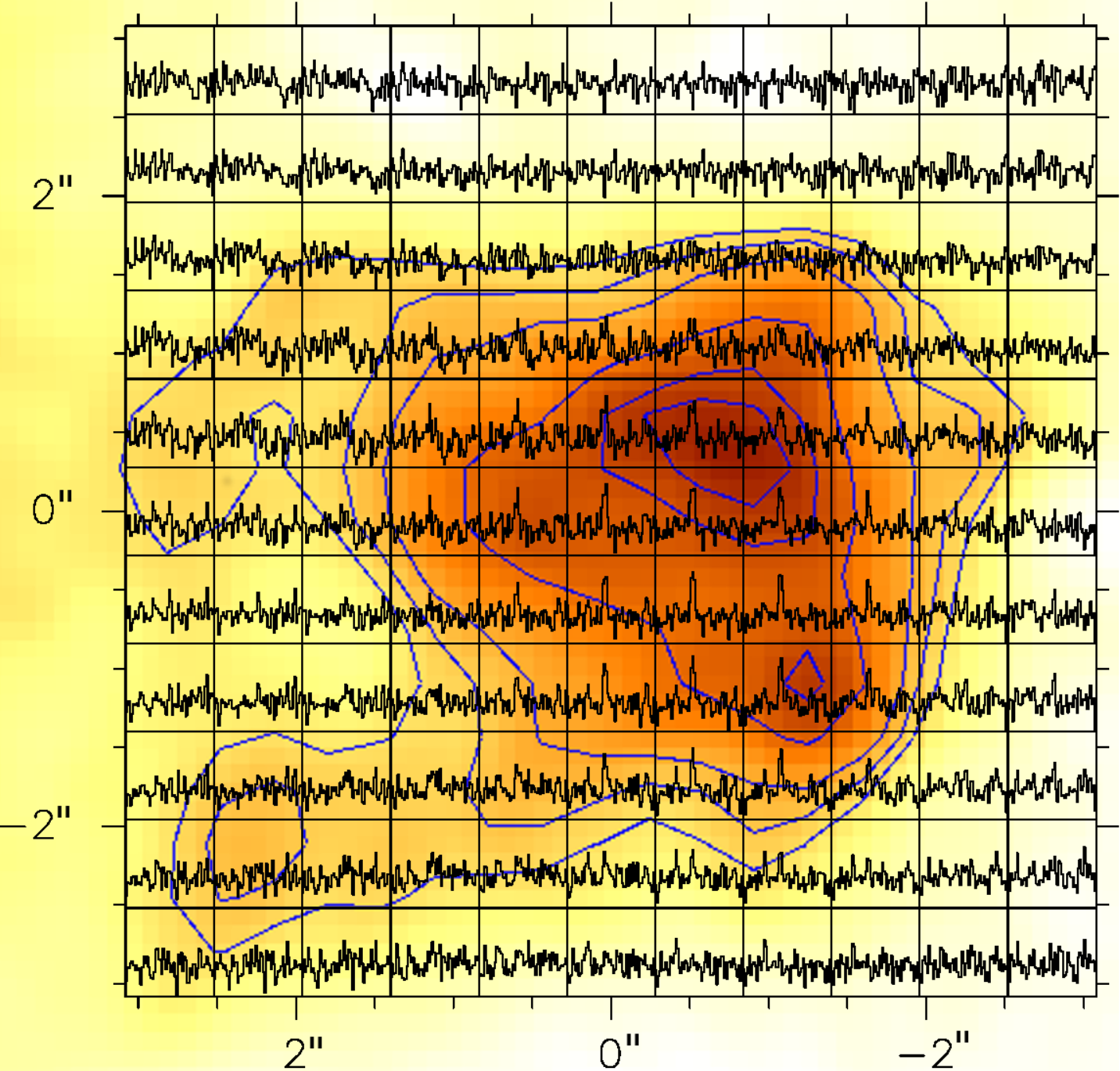}
\caption{Smoothed Magellan telescope IMACS $g$-band image of {\emph{the 
  patch}}. The contours at arbitrary levels
  follow the drop in the relative optical surface brightness.
  Overlaid are ALMA \twco (2--1) spectra smoothed by a
  2\arcsec\ Gaussian in the direction of {\emph{the patch}}. The spectra
  are twice Hanning-smoothed.}
  \label{Patch}
 \end{figure}

\begin{figure}
\centering
\includegraphics[width=8.8cm]{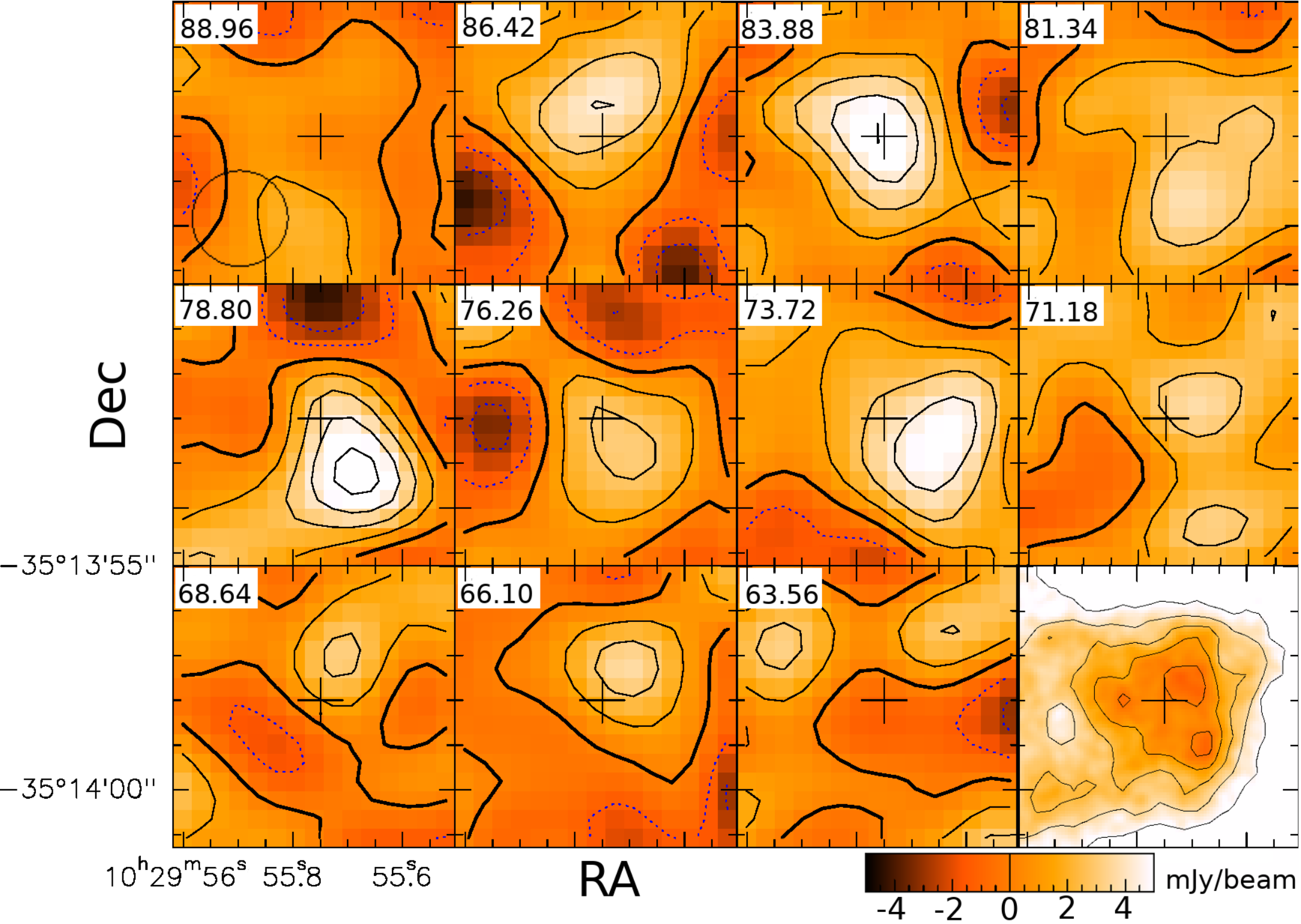}
\caption{Channel map of once-Hanning-smoothed spectra in the
  direction of {\emph{the patch}}. The velocity $V_{\rm r,hel} -3800$
  \,\kmps\ is shown in the upper-left corner of each panel. The cross
  marks the ALMA field phase centre, and the ALMA smoothed 2\farcs2 by
  2\farcs12 beam is shown in the panel in the upper-left. The
    contours are from 0 mJy/beam (thick contour) in one $\sigma$\ 
    step of a 1.4 mJy/beam.  Magellan Clay telescope MagIC $B$\ band
  image of {\emph {the patch}} is shown in the lower-right, where the
  contours at arbitrary levels follow the drop in the relative optical
  surface brightness.}
  \label{channel}
 \end{figure}

\section{Results} \label{res}

No line was detected above two sigma levels in either of the spectral
windows observed at APEX. {\emph{The patch}} was therefore re-observed
with ALMA, this time also including the intergalactic velocities.  {
  The three ALMA images were smoothed to a 2\farcs2 by 2\farcs12 beam
  and twice Hanning-smoothed to a channel width of 5.13\,\kmps\  in
  velocity to improve the noise level.  No signal above the noise was
  detected in the ALMA images centred on Galactic and intergalactic
  velocities.  A weak line was detected in the direction of {\emph{the
    patch}} in the spectral window centred on extragalactic
  velocities. The line intensity and line width from a Gaussian fit to
  the spectrum are 3.8\,mJy/beam$\pm$0.5\,mJy/beam (19\,mK$\pm$2.5\,mK in
  \Tmb\ scale) and 16.0\,\kmps$\pm$2.3\,\kmps, respectively. The LSR
  centre of line velocity is 3827\,\kmps$\pm$5.0\,\kmps,\ which, when
  expressed according to the optical redshift convention, corresponds
  to a {heliocentric} velocity of
  3878\,\kmps\ (Fig.\,\ref{Patchspec}).}

As a consistency check, we estimated the expected APEX
  CO\,(3--2) line strength based on our ALMA CO\,(2--1) detection.
  The dominating effect comes from the beam dilution: for a
  4\arcsec\ Gaussian distribution of the patch intensity, the dilution
  factor in an 18\arcsec\ beam is $\sim$20.  Adopting a CO\,(3--2) to
  CO\,(2--1) line intensity ratio of $\sim$0.6
  \citep{lampertietal2020}, the expected antenna temperature of the
  CO\,(3-2) line is $\sim$19\,mK/33 or $\sim$0.6\,mK. Compared with the
  15\,mK rms, this means that the sensitivity of our APEX observation
  was far below the detection limit required for {\emph{the patch}}.  
 
Twice-Hanning-smoothed spectra in the direction of {\emph{the patch}}
are shown {in Fig.\,\ref{Patch}} overlaid on the smoothed Magellan
telescope IMACS Sloan $g$-band image.  The contours trace the drop in relative optical
surface brightness of {\emph{the patch}}.  The image confirms that the
line emission originates in {\emph{the patch}} and that no other line
above the noise is seen outside {\emph{the patch}}.

A channel map of ALMA spectra once Hanning-smoothed to channel width
2.5\,\kmps\ in the direction of {\emph{the patch}} is shown in
Fig.\,\ref{channel}.  The optical attenuation in the field is shown in
the lower-right.  The two stronger attenuation maxima seen in the
inset in Fig.\,\ref{Magellan1} can be resolved in the channel map. The weaker
easternmost maximum is not seen, possibly due to the heavy spatial
smoothing.

The rms of the ALMA continuum image in the direction of {\emph{the patch}} is 
 20$\times 10^{-6}$\,Jy in a 1\arcsec\ beam. No sources above the noise
are detected in the image.

\begin{figure}[tbh]
\includegraphics[width=8.8cm]{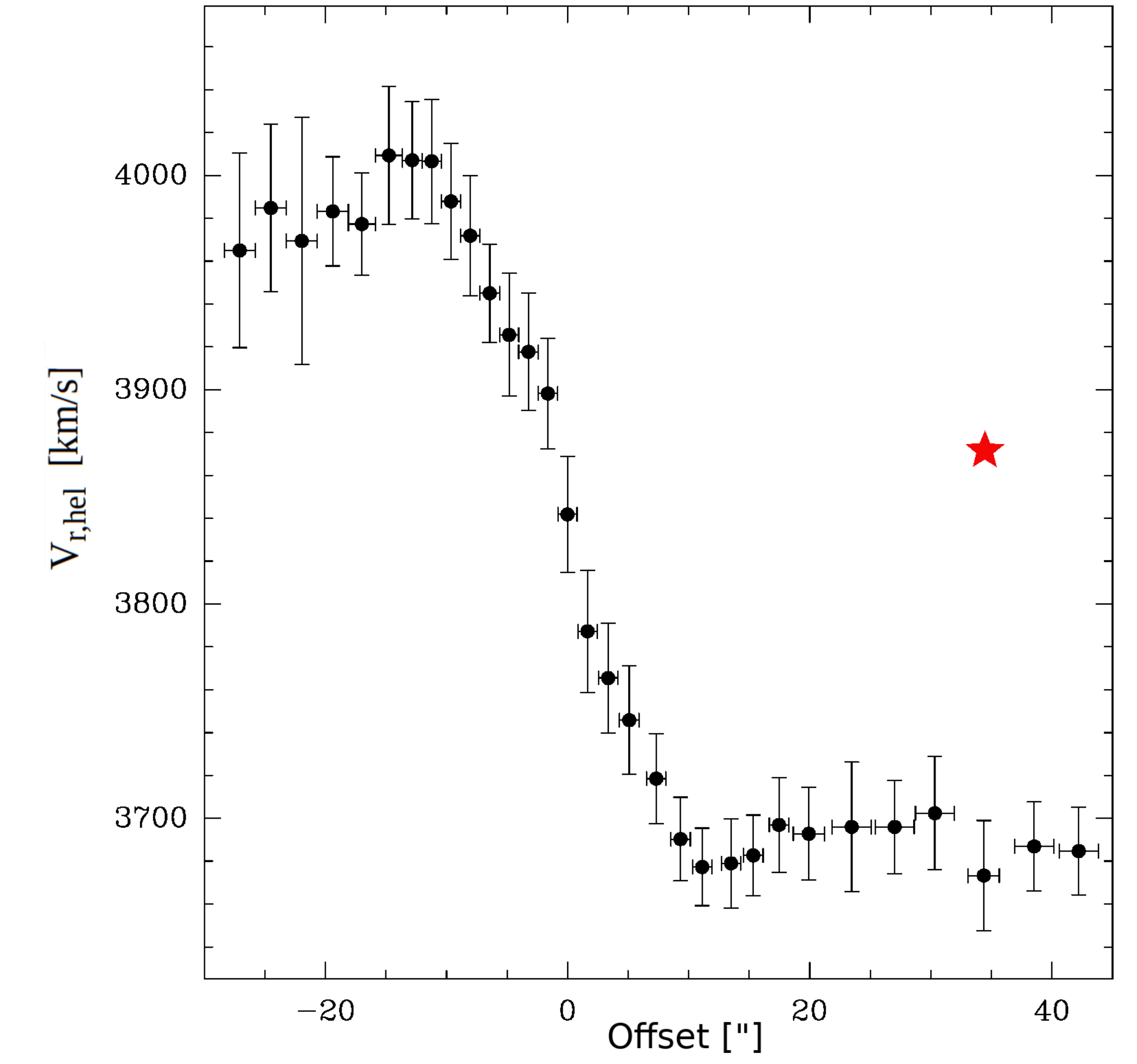}
\caption {Heliocentric radial velocities as measured from the
  GMOS long-slit spectrum.  The x-axis is the distance in
  arc seconds along the slit from the galaxy nucleus. {\emph{The patch}}
  is at an offset of 34\arcsec. The velocity of {\emph{the patch}} is shown with a
  red star.}
\label{Radvel}
\end{figure}

{The NGC\,3269 radial velocities along the GMOSS slit were measured from
  the extracted spectra with the cross-correlation procedure
  \citep{tonry79} as implemented in IRAF/\verb+fxcor+. The spectra
  were cross-correlated with 28 single stellar population synthesis
  models from \citet{vazdekis16}, with a 1--14 Gyr range of ages and
  $-1.3<$\,[M/H]\,$<$\,0.2 dex {metallicity.} The five results with
  the highest cross-correlation peaks were averaged for a final
  velocity per radial bin. The best fit SSPs were old (10--14 Gyr)
  and super solar ([M/H]=0.2) for the centre of the galaxy, and
  intermediate age (2 Gyr) and solar ([M/H]=0.0) for the disk. The
  resulting radial velocities along the slit are shown in Fig.  \ref
  {Radvel}.}

\section{Discussion} \label{discuss}

\subsection{Radial velocity of the patch}

The observed CO line velocity, $V_{\rm r,hel} = 3878\pm5.0$ \,\kmps,
agrees within the error limits with the radial velocity of the NGC\,3269
nucleus, $V_{\rm r,hel} = 3840\pm26$ \,\kmps\  as measured from the GMOS
spectrum.  This confirms the extragalactic nature of {\emph{the patch}}
and also shows that it is likely to be at the same distance as and
that associated with NGC\,3269.  The radial velocity of the NGC\,3269 disk in
the direction of {\emph {the patch}} has a radial velocity of $V_{\rm
  r,hel} = 3673\pm25$\,\kmps, that is,  $\sim$200\,\kmps \ smaller than that
of {\emph {the patch}}. This excludes the possibility that {\emph {the patch}}
could be a molecular cloud complex embedded in or participating alongside
the rotation of the galactic disk. Depending on its tangential
velocity component, {\emph {the patch}} could be falling into, orbiting
around, or passing by the galaxy NGC\,3269.

At the Hubble distance of NGC\,3269, 50.7\,Mpc, the 4\arcsec\ diameter of {\emph
  {the patch}} corresponds to nearly 1kpc.  The $\sim$1\arcsec\ size
attenuation maxima seen in Fig.\,\ref {Magellan1} inset have extensions
of $\sim$200\,pc each; {\emph {the patch}} is thus a large molecular cloud
complex.

 Given the substantial maximum optical attenuation of $A'_B \sim$1\fm0
 in {\emph {the patch}} \citep{dirschetal_2005}, the low intensity of the
 observed \twco~(2--1) line {of $\sim$0.3\,K\,\kmps\, as compared to
   $\ga$10 times larger values in Galactic molecular clouds with
   similar extinction,} is striking.  Possible explanations are that
 either the metallicity or the beam filling {of the molecular gas
   component in {\emph {the patch}} is extremely low. Even below the
   $\sim$1\arcsec\ clumping, seen in the optical image
   (Fig.~\ref{Magellan1}, inset), {\emph {the patch}} could consist of a
   large number of clouds with apparent diameters of only a fraction
   of an arcsecond, explaining the small beam filling.

\subsection{Mass of the patch}

The mass of {\emph {the patch}} can be estimated from the $^{12}$CO\,(2-1)
emission.  The average CO line integral $W_{CO} = \int{T dV}$ over
velocities 3863\,\kmps\ to 3894\,\kmps\ in an 22.6 arcsec$^2$\ area
where the line is detected (inside offsets -2\farcs5 to 2\farcs5 in RA
and -2\farcs5 to 2\farcs0 in DEC in Fig.~\ref{Patch}), is
0.92\,K\,\kmps\,arcsec$^{-2}$.  Adopting the \citet {herreraetal2020}
conversion factor $\alpha_{CO}$\,(2--1) of
6.2\,\Msun\,pc$^{-2}$\,(K\,\kmps)$^{-1}$, we obtain {\emph {the patch}}
mass of (d/50.7\,Mpc)$^2\times1.4\times 10^6$\,\Msun,\ where
d is the true distance to NGC\,3269.

\subsection{Mass of the dust}

The mass of dust in {\emph {the patch}} can be estimated by using the
attenuation\footnote{We emphasise the difference between the concepts
  of {\emph{extinction}} for point sources and {\emph{attenuation}} as
  obtained from dimming of background surface brightness that passes
  through a foreground dust screen.  The intensity ratio between the
  patch and the background gives the attenuation, which in magnitudes
  is expressed as \newline $A'= -2.5{\rm log}(I({\rm patch})/I({\rm background}))$.}
of the background light passing through {\emph {the patch}}.  The GMOS
spectrum of the surface brightness distribution across {\emph {the patch}}
was used to measure the attenuation over the wavelength range of
$\sim$\,4000\,\AA--\,6800\,\AA.  The spectrum was re-binned to
0\farcs64 in spatial coordinate along the slit, and the attenuation was
estimated by adopting a background intensity level 4\arcsec\ southwest of
{\emph {the patch}}. The NGC\,3269 disk surface
brightness in this position is similar to the background intensity in
the \citet {dirschetal_2005} virtual slit. It is assumed that the
spectral energy distribution in the off position is the same as that
behind {\emph {the patch}}. The detector gaps, bad pixels, and (because of
the low S/N) the wavelengths below 4000\,\AA\ were masked. The
resulting wavelength-dependent attenuation, after re-binning by 20 in
wavelength, is shown in Fig.\,\ref{Extinction} for a 0\farcs64 broad
slice at the position of the maximum attenuation. The blue line is a
third-degree polynomial fit to the data. The red and the dashed
green lines show the extinction wavelength dependence according to
\citet {Cardellietal1989} corresponding to A$_v$/E(B-V) values of 3.1
and 1.5. Our attenuation curve is seen to closely follow the latter
extinction curve.

  \begin{figure}[tbh]
\includegraphics[width=8.8cm]{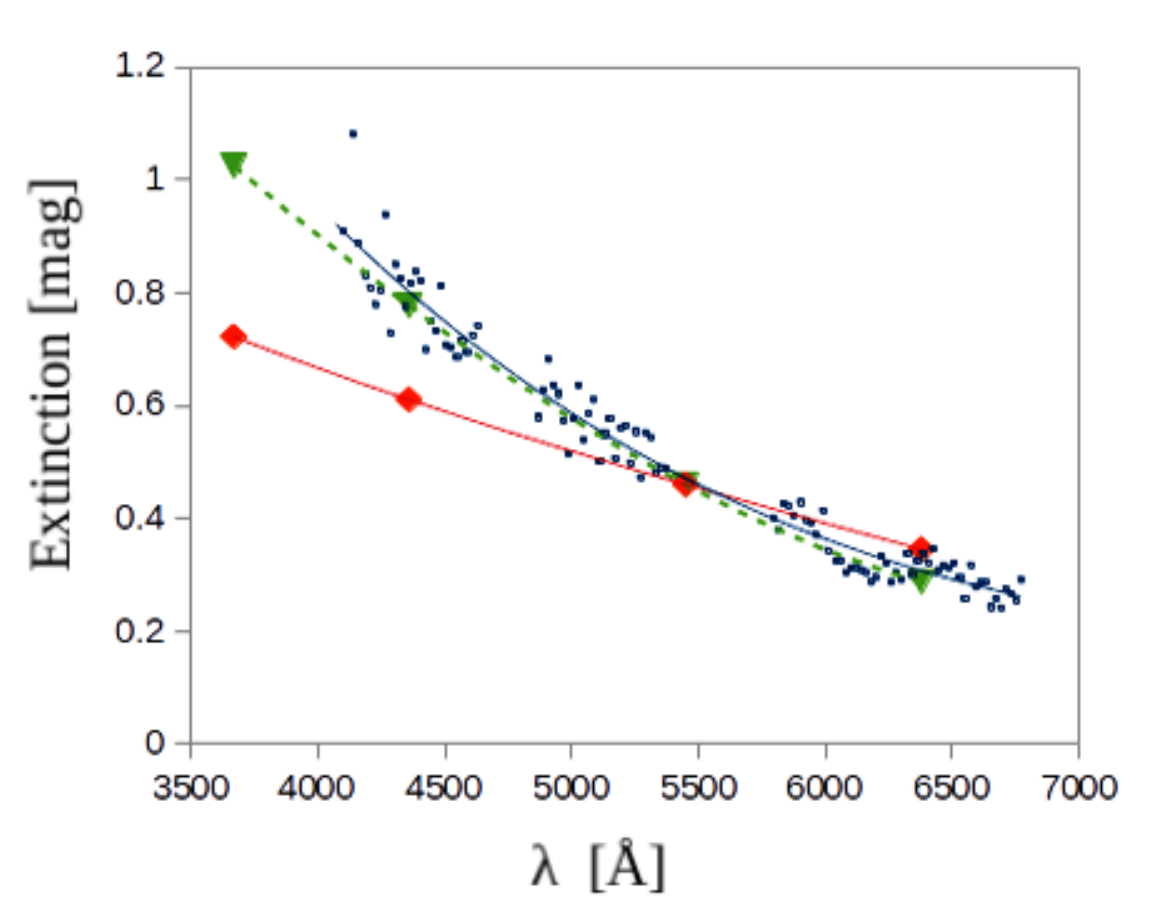}
\caption {Attenuation as calculated for a 0\farcs64 broad slice
  of GMOS spectrum in the position of the maximum attenuation of {\emph
    {the patch}}. GMOS pixels have been binned by 20 in the
  wavelength.  The blue line is a fit to the data. Extinction
    according to \citet{Cardellietal1989} is shown for A$_v$/E(B-V)
    values of 1.5 (green triangles, dashed line) and 3.1 (red
    diamonds, continuous line), respectively. }
\label{Extinction}
\end{figure}

\citet {dirschetal_2005} estimated the maximum
and the mean B-band attenuation in a 4\arcsec\ diameter area in {\emph
  {the patch}} to be $A'_B\sim$1\fm0 and 0\fm4, respectively.
Their maximum attenuation is somewhat higher than the value $A'_B$\,$
\approx0\fm8\pm$0\fm05 obtained from the GMOS spectrum. This
difference is possibly due to the positioning of the GMOS spectrum
slit that does not pass exactly through the densest part of the patch,
and also to the selected off position. Had the off position been chosen
northeast of {\emph {the patch}}, the attenuation would be 10\% lower, but the
wavelength dependence of the attenuation would not change.
Our $A'_B$/($A'_B-A'_R$) ratio ($1.6\pm$0.11) is the same as ($1.6\pm0.25$) 
reported in \citet {dirschetal_2005} and it is,
again, substantially lower than the corresponding value for the 
mean Milky Way extinction curve for point sources, A$_B$/(A$_B$-A$_R$)=2.3.}

{The background surface brightness attenuation in a clumpy foreground
  screen has been modelled by \citet{natta_panagia_1984}.  Clumping
  causes the reddening curve to flatten, meaning $R'_V$ and
  $A'_B/(A'_B - A'_R)$ increase as compared to a homogeneous
  screen. Thus, clumping cannot explain the steeper wavelength
  dependence as observed for {\emph {the patch}}.  In several cases, dust
  lanes or patches have been observed in early-type galaxies with
  $R'_V$ values as low as 1.9 \citep
  [\object {IC\,4320},][]{warren-smith_berry_1983} and 2.15 \citep
  [\object {NGC\,5626},][]{Goudfrooij_etal_1994}.  Such cases have been
  interpreted as evidence for dust-grain size distributions with mean
  and maximum grain sizes smaller than those for the mean Milky Way
  extinction curve (with $R_V = 3.1)$.} 

 The dust mass for lanes and patches in galaxies has frequently been estimated 
using the following formula:
\[ M_d = \Sigma \langle A'_B\rangle \Gamma^{-1}_B,\]
where $\langle A'_B\rangle$ is the mean attenuation as measured over
the object's area of $\Sigma$ \citep{sadler_gerhard_1985,
  vandokkum_franx_1995}.  For the Milky Way's mean dust  properties,
the value of the mass absorption coefficient $\Gamma_B$ is $\Gamma= 4\times
10^4$mag~cm$^2$g$^{-1} \sim 8 \times
10^{-6}$~mag~kpc$^2$\,\Msun$^{-1}$.  Adopting $ \langle A'_B\rangle =
0\fm4$\  as the mean attenuation over the area of $\Sigma = 4\arcsec\,
\times 4\arcsec\,$ of {\emph {the patch}}, we obtain the following for its {dust mass:
 $M_d = 4.8\times 10^4(d/50.7$\,Mpc)$^2$\,\Msun.}

\citet {Goudfrooij_etal_1994} presented a method to estimate the
value of $\Gamma_B$ for extinction (attenuation) curves with $R'_V$
values other than 3.1.  For NGC\,5626 and IC\,4320, which have the
smallest values of $R'_V$ in their sample (2.15 and 2.08), they found
that the dust masses as obtained from the above formula have to be
multiplied by a factor of $\sim$0.66. In the case of {\emph {the patch}}
with $R'_V = 1.5$, we estimate that a correction factor of $\sim$0.5
could be appropriate. {Thus the dust mass estimate for the patch
  becomes $M_d \sim 2.6 \times
  10^4(d/50.7\,\mathrm{Mpc})^2$\,\Msun,\ which corresponds to a
  molecular-gas-to-dust ratio of $\sim$55. If at the distance of
  24.3\,Mpc as implied by the Tully-Fisher argument (see
  Sect. \ref{origin}), the mass of the dust and molecular gas would be $\sim 6
  \times 10^3$\,\Msun\ and $\sim0.3 \times 10^6$\,\Msun, respectively,
  and the {molecular-}gas-to-dust ratio would not change.}

The attenuation of
  the patch caused by the dust is modest and it is very probable that
  {\emph {the patch}} also contains a large amount of atomic hydrogen.
With an amount similar to the molecular hydrogen, the gas-to-dust mass
would be close to the Milky Way standard value of ~100-150. The NGC\,3269 atomic
hydrogen mass upper limit for an assumed 200\,\kmps\ line width 
is $\sim$ $10^8$\,\Msun\ \citep{barneswebster2001}. Even for a much
narrower line, the \ion{H}{I}  mass upper limit remains much above the
molecular gas mass of   {\emph {the patch}} and the \ion{H}{I}
gas could thus easily respond to the need to make the gas-to-dust mass
similar  to the Milky Way standard value.

 In a surface brightness distribution like the face of a galaxy, 
 instrumental and atmospheric 'PSF blurring' tends to fill
in {\emph small size} (1\arcsec--\,4\arcsec) depressions or
'holes', such as  {\emph the patch} and its sub-structures.  
Light from the surrounding region of higher surface 
brightness is being poured into 'the dark hole'. Given the
instrumental+atmospheric PSF this effect could, at least
approximately, be corrected for as has been discussed by
\citet{McCaughrean1996}. They analysed the effect in the
case of their $HST$ imaging of circumstellar disks, seen silhouetted as
dark markings against the bright background of the Orion Nebula. They
found that the 'PSF blurring' became an important correction for their
$\la$0.5\arcsec sized objects.  Given the substantially broader core
and wider wings of the PSFs of ground-based telescopes, the blurring
effect is already important for source structures of 1\arcsec -
4\arcsec , as in {\emph {the patch}}.
Correction of this effect has not been attempted for {\emph {the patch}}. 
It would have had the effect of increasing the attenuation 
and thereby the dust mass estimate.

\subsection{Origin of the patch} \label{origin}

The confirmation of {\emph {the patch}} as a projected Galactic dust cloud
would have been an intriguing finding. However, now that its
extragalactic nature is obvious, the resulting questions are no less
intriguing. Dust in early-type galaxies is frequent, but an isolated
large dust patch in a galaxy without young stellar populations is, to
our knowledge, very rare (if not unique) in the literature. A
morphologically similar patch in \object {NGC\,3923} was found by
\citet {Sikkemaetal2007}.  However, NGC\,3923 is a prominent shell galaxy
with obvious infall of dwarf galaxies, and there is more dust beside that
patch. Even in this case,  \citet {Sikkemaetal2007} favour an internal origin.
This may be the case in NGC\,3269 as
well. NGC\,3269 is a transition object between a grand design spiral
and an S0, so isolated dust suggests relation to the
quenching of star formation, (i.e. the process of transforming a star-forming
spiral galaxy into a non-star-forming S0). In that way, one
could understand the dust in NGC\,3269 as a debris of the former
gas/dust component of its disk.

The galaxy NGC\,3269 looks tidally undisturbed at first
sight. However, the grand spiral design by itself may be a sign of
previous tidal influence \citep
[e.g.][]{struck2012,semczuketal2016}. Other traces of a tidal history
were noted by \citet[][see the inset of their Fig.1]{dirschetal_2005},
in particular an arc-like structure extending towards the
southeast. Because this arc is much more visible in the Washington
C-filter than in the faster R-filter, one suspects that the enhanced
brightness is due to the very strong \ion{O}{III}-line at 3727\AA\ in
\ion{H}{ii}-regions that falls into the C-filter. It is a clear sign
of star formation, but with our present data, we cannot further
constrain possible tidal effects.

A viable mechanism of removing gas from galaxies is ram-pressure
stripping by a passage through a hot ambient medium
(e.g. \citet{abramsonetal2011,jaffeetal2017,crameretal2019}).  The
obvious candidate for this process is the extended hot halo of \object
{NGC\,3268}, whose projected distance is only 60 kpc, but its radial
velocity is 2700\,\kmps, 1000\,\kmps\ less than that of NGC\,3269.
By its position, the arc could principally indicate the wake of the
passage. Assuming this, NGC\,3269 would be in the foreground of
NGC\,3268, the closest distance yet to come. Otherwise, a passage that
close to NGC\,3268 would have left a more disturbed structure of both
NGC\,3268 and NGC\,3269. That contradicts the argument of
\citet{casorichtler2015}, who put NGC\,3269 far in the background. They
interpret the radial velocity difference of 1000\,\kmps\ between
NGC\,3268 and NGC\,3269 as a difference of the recession velocities
rather than of the Doppler velocities.  A further striking feature of
that galaxy group around NGC\,3268 is that
the other two S0-galaxies, \object{NGC\,3271} and \object{NGC\,3267},
show radial velocities of 3804\,\kmps\ and 3709\,\kmps, respectively, which is almost
identical to that of NGC\,3269.  This group would then be at a
distance of about 50 Mpc, and tidal interactions between these
galaxies would be possible, but ram-pressure effects would lose their
plausibility.  The Antlia galaxy cluster in this case would be an
effect of looking along a filament of galaxies, which results in a
superposition of Doppler and recession velocities.

The clear key to this problem lies in the precise
individual distances to these three galaxies, which are currently not
available. The distance to NGC\,3268 is, however, well established.

We can make a rough estimate by assuming that NGC\,3269, like all disk
galaxies, is on the baryonic Tully-Fisher relation.  We adopt
the relation $M_{bar} = 50 \times v_f^4$ \citep{mcgaugh2005} and assume
that the observed radial velocities are the circular velocities, which may be
wrong, if tidal effects are important. The
radial velocity difference between the centre and the constant level
in the disk is 150\,\kmps,\ the angle of inclination is 27.5\degr, the
position angle of the slit with respect to the major axis is 10\degr:
this gives 175\,\kmps\ and $M_{bar} = 4.7\times10^{10} M_\odot$. For an old population, we
adopt  $M/L =5.0$. In reality, it will be
somewhat smaller because of the intermediate-age disk. This gives
$9.4\times10^9$ solar luminosities, or $M_R = -20\fm7$. The apparent
R-magnitude is 11\fm73. With $A_R = 0\fm2$, one has m-M = 31\fm9 or
24.3\,Mpc (i.e. in the foreground of NGC\,3268 by 10\,Mpc).  The
uncertainties are unknown and probably large, but a Hubble distance of 50\,Mpc
is difficult to justify. This raises the problem of explaining the
radial velocities of the three S0s. If they are a group, as the very
similar radial velocities suggest, the velocity difference to
NGC\,3268 cannot be due to the gravitational attraction of the Antlia
cluster, but must be attributed to a large-scale structure.

\section{Conclusions} \label{conclusions}

The S0 galaxy NGC\,3269 is devoid of gas, but hosts a dust patch,
which has previously been suspected as a Galactic tiny dust cloud,
projected onto NGC\,3269. While APEX observations failed to detect the
dust patch in the \twco (3--2) line, faint \twco (2--1) emission
associated with the dust was detected with ALMA.  {\emph {The patch}}
radial CO velocity, $V_{\rm r,hel} = 3878\pm5.0$\,\kmps, places it
clearly at the galaxy's distance and matches the radial velocity of
the galaxy's nucleus. The patch deviates from the radial velocity
curve of the disk of NGC\,3269, as shown through Gemini GMOSS
long-slit spectroscopy.
The optical attenuation as estimated  from the long-slit
spectrum  confirms the steepness of the reddening law reported by
\citet{dirschetal_2005}.
The gas mass estimate from the CO emission is approximately
1.4$\times10^6(d/50.7\,\mathrm{Mpc})^2$\,\Msun.
The origin of the dust patch remains, however, elusive. A plausible
speculation is that the gas/dust cloud is debris from the previous
gas/dust component of the disk of NGC\,3269. A possible process is
ram-pressure stripping by the hot halo of the neighbouring galaxy, NGC\,3268,
which requires NGC\,3269 to be in the foreground of NGC\,3268.
This revives the problem of the nature of the radial velocities
of NGC\,3269 and its neighbouring S0s, NGC\,3267 and NGC\,3271: are these recession
velocities or Doppler velocities? Robust distances are needed to solve
this problem.

\begin{acknowledgements}
  TR acknowledges support  from  the BASAL Centro de
 Astrof\'{\i}sica y Tecnologias Afines (CATA) PFB-06/2007.
      Based partly on observations obtained at the international
      Gemini Observatory, a program of NSF’s NOIRLab, which is managed
      by the Association of Universities for Research in Astronomy
      (AURA) under a cooperative agreement with the National Science
      Foundation. on behalf of the Gemini Observatory partnership: the
      National Science Foundation (United States), National Research
      Council (Canada), Agencia Nacional de Investigaci\'{o}n y
      Desarrollo (Chile), Ministerio de Ciencia, Tecnolog\'{i}a e
      Innovaci\'{o}n (Argentina), Minist\'{e}rio da Ci\^{e}ncia,
      Tecnologia, Inova\c{c}\~{o}es e Comunica\c{c}\~{o}es (Brazil),
      and Korea Astronomy and Space Science Institute (Republic of
      Korea) and partly on data acquired with the Atacama Pathfinder
      EXperiment (APEX) (project O-079F-9321A). APEX is a
      collaboration between the Max–Planck–Institut für
      Radioastronomie, the European Southern Observatory, and the
      Onsala Space Observatory.  This paper includes data gathered
      with the 6.5 meter Magellan Telescopes located at Las Campanas
      Observatory, Chile obtained via CNTAC program CN2018A-83\\
      This paper makes use of the following ALMA data:
      ADS/JAO.ALMA\#2017.1.00066.S. ALMA is a partnership of ESO
      (representing its member states), NSF (USA) and NINS (Japan),
      together with NRC (Canada), MOST and ASIAA (Taiwan), and KASI
      (Republic of Korea), in cooperation with the Republic of
      Chile. The Joint ALMA Observatory is operated by ESO, AUI/NRAO
      and NAOJ.
\end{acknowledgements}
\bibliographystyle{aa}
\bibliography{38994_Haikalaetal.bib}
\end{document}